%Paper: hep-ph/9308277
%From: anelson@UCSD.EDU
%Date: Sun, 15 Aug 1993 15:56:28 -0700

\input harvmac
%%%%%%%%%%%%%%%%%%%%%%%%%%%%%%%%%%%%%%%%%%%%%%%%%%%%%%%%%%%%%%%%%%%%%%
%
%  UCSD macros to overwrite some of the definitions in harvmac.tex
%  (include after harvmac.tex)
%  last modified 4/92
%
%%%%%%%%%%%%%%%%%%%%%%%%%%%%%%%%%%%%%%%%%%%%%%%%%%%%%%%%%%%%%%%%%%%%%%%
%
% modify the output routine for the little format
%
\ifx\answ\bigans
\else
\output={
  \almostshipout{\leftline{\vbox{\pagebody\makefootline}}}\advancepageno
}
\fi
%
%
% address
%
\def\mayer{\vbox{\sl\centerline{Department of Physics 0319}%
\centerline{University of California, San Diego}
\centerline{9500 Gilman Drive}
\centerline{La Jolla, CA 92093-0319}}}
\def\title#1{\nopagenumbers\hsize=\hsbody%
\centerline{\titlefont #1} \tenpoint \vskip .5in\pageno=0}
%
%
% grant numbers
%
\def\doe{\#DOE-FG03-90ER40546}
\def\tnlrc{\#RGFY93-206}
%
% preprint number
%
\def\UCSD#1#2{\noindent#1\hfill #2%
\bigskip\supereject\global\hsize=\hsbody%
\footline={\hss\tenrm\folio\hss}}% restores pagenumbers
%
% abstract
%
\def\abstract#1{\centerline{\bf Abstract}\nobreak\medskip\nobreak\par #1}
%
%
% titlefont
%
%
\edef\tfontsize{ scaled\magstep3}
 \tfontsize  \tfontsize
 \tfontsize \font\titlei=cmmi10 \tfontsize
\font\titleis=cmmi7 \tfontsize \font\titleiss=cmmi5 \tfontsize
\font\titlesy=cmsy10 \tfontsize \font\titlesys=cmsy7 \tfontsize
\font\titlesyss=cmsy5 \tfontsize  \tfontsize
\skewchar\titlei='177 \skewchar\titleis='177 \skewchar\titleiss='177
\skewchar\titlesy='60 \skewchar\titlesys='60 \skewchar\titlesyss='60
%
%\def\titlefont{\def\rm{\fam0\titlerm}% switch to title font
%\textfont0=\titlerm \scriptfont0=\titlerms \scriptscriptfont0=\titlermss
%\textfont1=\titlei \scriptfont1=\titleis \scriptscriptfont1=\titleiss
%\textfont2=\titlesy \scriptfont2=\titlesys \scriptscriptfont2=\titlesyss
%\textfont\itfam=\titleit \def\it{\fam\itfam\titleit}\rm}
%
%
% math symbols
%
%---------------------------------------------------------------------
%
%
%
% space and backspace in l mode
%
\def\lspace{\ifx\answ\bigans{}\else\qquad\fi}
\def\lbspace{\ifx\answ\bigans{}\else\hskip-.2in\fi} % $$\lbspace...$$
%
%
%     curly letters
%
   %curly letters

  \def\CO{{\cal O}}

%
%
%
%     derivatives
%
%

%

\def\bar#1{\overline{#1}}
\def\vev#1{\left\langle #1 \right\rangle}

\def\half{{\textstyle{1\over2}}} %puts a small half in a displayed eqn
\def\frac#1#2{{\textstyle{#1\over #2}}} %puts a small fraction
%in a displayed eqn
%
%
%     various math operators
%
%

\def\GeV{{\rm GeV}}

\def\hc{\rm h.c.}
%
%
%
%       relations
%
\def\ltap{\ \raise.3ex\hbox{$<$\kern-.75em\lower1ex\hbox{$\sim$}}\ }
\def\gtap{\ \raise.3ex\hbox{$>$\kern-.75em\lower1ex\hbox{$\sim$}}\ }
\def\gl{\ \raise.5ex\hbox{$>$}\kern-.8em\lower.5ex\hbox{$<$}\ }
\def\roughly#1{\raise.3ex\hbox{$#1$\kern-.75em\lower1ex\hbox{$\sim$}}}
%
%
%       This defines et al., i.e., e.g., cf., etc.

%
\def\np#1#2#3{{Nucl. Phys. } B{#1} (#2) #3}
\def\pl#1#2#3{{Phys. Lett. } {#1}B (#2) #3}
\def\prl#1#2#3{{Phys. Rev. Lett. } {#1} (#2) #3}
\def\physrev#1#2#3{{Phys. Rev. } {#1} (#2) #3}

\def\prep#1#2#3{{Phys. Rep. } {#1} (#2) #3}

\def\frac#1#2{{\textstyle{#1 \over #2}}}

\def\[{\left[}
\def\]{\right]}
\def\({\left(}
\def\){\right)}

\def\mayer{\vbox{\sl\centerline{Department of Physics}
\centerline{9500 Gilman Drive 0319}
\centerline{University of California, San Diego}
\centerline{La Jolla, CA 92093-0319}}}
\def\tb{$\tan\beta$}
\noblackbox
%%% start paper
\centerline{{\titlefont{ Naturally Large Tan $\beta$
}}\footnote{$^*$}{This work is supported in part by funds provided
by the U. S. Department of Energy (D.O.E.) under contracts
\#DE-AC02-76ER03069 and \doe\
and in part by the Texas National Research Laboratory Commission
under grants \#RGFY92C6 and \tnlrc. }}
\def\awards{National Science Foundation Young Investigator Award.\hfill\break
Department of Energy Outstanding Junior Investigator Award.\hfill
}
\def\sloan{Alfred~P.~Sloan Foundation Research Fellowship.\hfill}

\vskip.3truein
\centerline{Ann E. Nelson\footnote{$^\#$  }{\sloan}}
\medskip
\mayer
\bigskip
\centerline{
Lisa Randall$^\#$\footnote{$^{\ddag}$}{\awards} }
\medskip
\vbox{\sl\centerline{ Center for Theoretical Physics}
\centerline{\sl Laboratory for Nuclear Science and Department of
Physics}
\centerline{\sl Massachusetts Institute of Technology}
\centerline{\sl Cambridge, MA 02139}}
\bigskip
\vfill
\abstract{We show that if there are only two Higgs doublets in the
supersymmetric standard model,  large \tb\ requires a fine tuning in the
parameters of the Lagrangian of order (1/\tb), which cannot be explained by any
approximate symmetry.
With an extended Higgs sector, large \tb\ can be natural.  We give
an explicit example
with four doublets in which it is possible to achieve large \tb\
as a result of an approximate symmetry, without any light superpartners. The
approximate symmetry can be extended to explain all the hierarchies in the
quark
mass matrix. }\vfill
\UCSD{\hbox{UCSD/PTH 93-24, MIT-CTP-2230, hep-ph@xxx/9308277}}{July 1993}

\newsec{Introduction}

The large ratio between the top and bottom quark masses is one of many puzzling
hierarchies in the standard model. One possible explanation
for this hierarchy arises in supersymmetric theories, which
require at least two Higgs doublets, one which couples to up
type quarks and one   to down quarks. In these theories,
the ratio of masses could conceivably arise entirely from a large
value of $\tan\beta$, the ratio of the vacuum expectation values
of the up and down type Higgs fields \nref\als{ B. Ananthanarayan,   G.
Lazarides,  Q. Shafi, \physrev{41}{1991}{1613},
\pl{300}{1993}{245}}\nref\hall{L.J. Hall, R. Rattazzi,
U. Sarid, Berkeley preprint UCB-PTH-93/15, (1993)} \refs{\als, \hall}.

However, a large value of $\tan\beta$ is also puzzling,
as one would expect all vacuum expectation values to be roughly
equal and set by the weak scale.    We argue
that any such model with a minimal Higgs sector requires
at least one fine
tuning of the parameters of the potential of order 1/\tb\ in order to achieve
electroweak symmetry breaking at the observed scale while avoiding overly light
charginos.
In a recent paper, Hall, Rattazzi
and Sarid (HRS) \hall\ explain this large ratio entirely through
approximate symmetries.
Consistent with our claim, the approximate symmetries they introduce reduce the
number of fine tunings but do not entirely eliminate them.

The fine tuning  {\it can} be eliminated  at the expense
of introducing additional fields. We give an illustrative example
of a  model
with four Higgs doublets  which naturally
has large \tb\ without any light fields, indicating that
the predictions of large \tb\ scenarios are very model dependent.

\newsec{Large \tb\ with a minimal Higgs sector}

In this section, we show that given the constraints on charginos
and higgsinos, large \tb\ necessarily implies a large hierarchy
(of order \tb) between the various mass squared parameters of the potential.
Approximate symmetries \hall\ can enforce the relatively small size of some
parameters, {\it but not all of them}. There are several different philosophies
as too what constitutes an acceptable hierarchy between the various
parameters.    Our naturalness  criterion is that  unless constrained by
additional approximate symmetries, all mass parameters are about the same
size, and all dimensionless numbers
are of order one. We then  find that \tb\ should also be of order one in a
phenomenologically acceptable supersymmetric model with only
two Higgs doublets.

\def\mup{m_U}
\def\md{m_D}
\def\hu{H_U}
\def\hd{H_D}
\def\er{$ \epsilon_{{}_R}$}
\def\ep{${\epsilon_{{}_{PQ}}}$}

Let us assume the standard scalar potential for the two Higgs
fields, $H_u$ and $H_d$.
\eqn\potential{\mup^2 \hu^2+\md^2\hd^2+ B \mu \left(\hu \hd+ \hc\right)
+\frac{g^2+{g'}^2}{8} \left( |\hu|^2-|\hd|^2\right)^2 \ ,} where $\mu$ is
the higgsino mass parameter in the superpotential.  If the electroweak scale is
to arise naturally, one would expect the parameters $\mup^2$ and $\md^2$ to
be of
order $m_Z^2$. Tan$\beta$ can be computed from this potential to be
\eqn\tanbeta{ \half\sin 2 \beta=-{\mu B\over \mup^2+\md^2}\approx {1\over
\tan\beta}\ .} Thus large \tb\ requires the parameter $\mu B$ to be much
smaller than $\mup^2+\md^2$. Naturalness requires that a parameter can only be
small if a symmetry is restored in the limit that the parameter goes to zero.
As
pointed out by HRS,  $B$ can naturally be made small by an approximate R
symmetry, while $\mu$ could be small because of an approximate Peccei-Quinn
(PQ) symmetry, which is taken to commute with supersymmetry. However in a two
Higgs doublet model, either of these symmetries results in a light chargino.
Let
the parameter which suppresses $R$ symmetry breaking terms be \er. The   PQ
symmetry is only broken by the   parameter  $\mu$. Then the chargino mass
matrix (in a basis where the first entry refers to a Higgsino and the second
entry to a chargino) takes the form
\eqn\massmatrix{\pmatrix{\mu&{g\over\sqrt{2}}\vev{\hu}\cr {g\over\sqrt 2}
\vev{\hd}&  \epsilon_R  M_S  } \ , } where $M_S$ is of order the supersymmetry
breaking scale. Now because $\vev{\hd}$ is assumed to be small, there will be a
light eigenvalue unless both $\mu$ and $\epsilon_R M_S$ are large. However a
light chargino is unacceptable. Because the lightest eigenstate must exceed 45
GeV,  if \tb\ is much larger than one we are forced to take both $\mu>85$
GeV and
$\epsilon_R M_s >85$ GeV. Now since $  B$ is of order  $\epsilon_R M_S$, while
$\mup^2+\md^2$ is naturally of  order $M_s^2$, eq.~\tanbeta\ requires that
\eqn\susyscale{M_s^2>{(85\; \GeV)^2 \tan\beta}\ .} Thus if there is an
approximate symmetry  forcing \tb\ to be large,   the supersymmetry scale is
also  forced to be much larger than the weak scale,  contrary to what  would be
true in a natural scenario.

The reason the minimal model is so constrained is readily understood
on the basis of chiral charge (or alternatively anomaly) assignments.
If both a PQ symmetry and $R$ symmetry are maintained (notice
the charges may be chosen so the vacuum expectation value of
$H_U$ breaks neither), the chiral charge assignments require
the presence of an additional massless charged fermion in addition
to the down quarks.

Let us our compare our result to the parameters in the paper of
HRS. In order to achieve a \tb\ as large as 50, they take $\md^2$ to be about
$(700\;\GeV)^2$ while both $B\mu$ and $\mup^2$ are of order $-m_Z^2$, a
factor of 50   smaller. While the relatively small size of $B\mu$ is enforced
by
a combination of PQ and $R$ symmetries, there is no symmetry which can suppress
$\mup^2$, which is nonetheless required to be a small negative number since
it sets the scale of the $W$ and $Z$ masses. Thus there is a   hierarchy in the
Higgs potential  of order \tb\ which is not enforced by any symmetry,
although it
can be arranged in a radiative symmetry breaking scenario by tuning the top
mass.
Our aim is to
generate a model with large \tb\  which is devoid of
fine tuning, which we have shown
is not possible in the context of the minimal Higgs sector,
given the light chargino constraint.
In the next section, we show  that  with an enlarged
Higgs sector, large \tb\ can be
naturally  arranged.

\newsec{A model with naturally large \tb}

The minimal extension of the Higgs sector is the addition of a gauge
singlet $S$.
However, one can easily see that this does not help make large \tb\ more
natural. With a gauge singlet one can add to the superpotential   a term
$\lambda S\hu\hd$, which   contributes to the Higgsino mass if $S$ gets a vev.
One then should also add to the potential \potential\ a term $B' \lambda S
\hu\hd$, where $B'$ is of order $M_S$ unless one imposes an approximate
R symmetry, in which case $B'\sim\epsilon_RM_S$. Now \tb\ is of order $(m_U^2
+\md^2)/(B'\lambda\vev S+B\mu)$, which is naturally large if either \er is
small or $\lambda\vev S$ and $\mu$ are small. With only two Higgs doublets, the
chargino mass matrix is now \eqn\massmatrixp{\pmatrix{\mu+\lambda\vev
S&{g\over\sqrt{2}}\vev{\hu}\cr {g\over\sqrt 2} \vev{\hd}& M_S   \epsilon_R } \
,
} which will  have a light eigenvalue if \tb\ is naturally
large,     and $\mup^2$ and $\md^2$ are both of order $M_S^2$. To avoid the
light chargino   would require a model in which the parameters providing
Higgsino masses are not suppressed and in which there is no approximate $R$
symmetry.
But then the model
is no more natural than  the minimal model. This
result can be easily understood by looking at the charged fermion   charge
assignments under the approximate chiral symmetries, and noting that with only
two Higgs doublets any symmetry which protects \tb\ also requires additional
light charged particles.

We therefore are led to the addition of another Higgs doublet.
Anomaly constraints require two new Higgs doublets. (This
would change the prediction of $\sin^2 \theta_W$ in a GUT
scenario unless we also add other particles such as a vector-like color
triplet).  We  add two additional Higgs particles, which we call
\def\hup{H_U'}
\def\hdp{H_D'}
$\hup$ and $\hdp$.  We assume \tb\ is
large; that is, there is  a large vacuum expectation
value  for $\hu$, and a small one for $\hd$. To suppress the term $\mu_{ud}
\hu\hd$, which would induce a large vev for $\hd$,  we   impose an
approximate PQ
symmetry  under which   $\hd$ has charge 1, the left handed down antiquarks
have
charge -1, and all other quarks and $\hu$ have charge  0.  (We do not wish to
impose an approximate R symmetry because this would lead to a  light chargino
mass.) Now the chargino mass matrix is
\eqn\newchmass{\pmatrix{\mu_{ud}&\mu_{ud'}  & {g\over\sqrt{2}}\vev{\hu}\cr
\mu_{u'd}&\mu_{u'd'}&{g\over\sqrt{2}}\vev{\hup}\cr {g\over\sqrt 2}
\vev{\hd}&{g\over\sqrt 2} \vev{\hdp}& \CO(M_S)  \cr   } \ , } which, since
$\mu_{ud}$ and $\vev{\hd}$ are assumed to be small, will have a light
eigenvalue
unless $\mu_{u'd}$ is reasonably large. Thus $\hup$ should have PQ charge -1,
and $\vev\hup$ should   be small, or $\mu_{u'd}$ will induce a large vev for
$\hd$. Now   we   need to determine the PQ charge of $\hdp$. This can be chosen
such that the PQ charge assignments of the charginos are vectorlike, so that no
chargino masses are suppressed by the approximate symmetry. Therefore we take
$\hdp$ to have a PQ charge of 0.  \foot{For nonzero choice of the charge of
$H_d'$, avoiding a light chargino would   require spontaneous breakdown of
the PQ
symmetry at the weak scale. Then  one would be forced to complicate the model
further in order to achieve a stable vaccuum with spontaneously broken PQ
symmetry, and there would also be   a light pseudo Goldstone boson.}  Then in
the limit that the PQ symmetry is exact, if   only $\hu$ has negative mass
squared,    only $\hu$ and $\hdp$ receive vevs, and the chargino   mass matrix
has the form \eqn\pqchmass{\pmatrix{0&\mu_{ud'}  &{g\over\sqrt{2}}\vev{\hu}\cr
\mu_{u'd}&0&0\cr  0&{g\over\sqrt 2} \vev{\hdp}& \CO(M_S)  \cr   } \ , } which
naturally has all eigenvalues of order the weak scale. In this limit there is
also no Goldstone boson, since the PQ symmetry is not spontaneously broken,
however the charge -1/3 quarks are massless. Allowing small explicit violation
of the PQ symmetry by an amount  \ep will give $\hd$ and $\hup$ small vevs and
the down type quarks small masses.  Note that the small explicit PQ violation
does not imply that there must be a light pseudo Goldstone boson, even though
it
induces small vevs for the scalars carrying nonzero PQ charges, since the size
of the vevs is no larger than the explicit symmetry breaking terms.

Thus by making a vectorial assignment of PQ charge on the charginos, and  by
avoiding an
$R$ symmetry, we have naturally large \tb\ without an additional light state
in the theory.

A potential problem with this  particular
model is that with extra Higgs doublets it is
possible to have more than one doublet couple to quarks with the same charge,
leading to the possibility of large flavor changing neutral currents (FCNC).
This does not happen in the limit that the PQ symmetry is exact, however
explicit PQ violation would allow FCNC. For instance the $\hdp$   coupling to
down type quarks could be suppressed by \ep. This does  not  provide sufficient
suppression of FCNC. The vev of  $\hd$ is also suppressed relative to
$\vev\hdp$ by an amount \ep, and so $\hd$ and $\hdp$ would make comparable
contributions to the down quark masses. In models with \tb\ of order one where
all Yukawa couplings to the quarks have the same approximate ``texture''
enforced
by approximate chiral symmetries, it is possible to have more than one Higgs
couple to each type of quarks without inducing unacceptably large FCNC
\ref\hallagain{A. Antaramian, L. J. Hall, A. Rasin
\prl{69}{1992}{1871};  L. Hall, S. Weinberg,
\physrev{48}{1993}{979}}. In our large \tb\ case however, the FCNC
in the down quark sector are enhanced relative to the \tb\ $\sim 1$ case
by a factor of  \tb, and would generally be too large.

Of course supersymmetric theories already have   potential
FCNC  which can arise from squark exchanges \ref\susyfcnc{e.g. H. P. Nilles,
\prep{110}{1984}{1}; H. Georgi,
\pl{169}{1984}{231}; L.J. Hall, V.A. Kostelecky,   S. Raby,
\np{267}{1986}{415}}. We will assume that
the problem of FCNC from the soft supersymmetry breaking terms has been solved
by some means \nref\sfcncsol{M. Dine, A.E. Nelson, \physrev{48}{1993}{1277};
M. Dine, A. Kagan, R. Leigh, UCSC preprint SCIPP 93/04
(1993); B. de Carlos, J.A. Casas, C. Mu\~noz, \pl{299}{1993}{234}; V.S.
Kaplunovsky, J. Louis, CERN preprint CERN-TH 6809/93 (1993); R. Barbieri, J.
Louis, M. Moretti, CERN preprint CERN-TH 6856/93 (1993) }\nref\nir{Y. Nir, N.
Seiberg, Rutgers preprint RU-93-16 (1993)} \refs{\sfcncsol, \nir}, but we still
need to avoid FCNC from Higgs exchange by finding some   way of further
suppressing the coupling of $\hdp$ to the down quark sector.

 One simple possibility utilizes the same
mechanism that Nir and Seiberg \nir\ used to obtain zeros in quark mass
matrices.
The small parameter \ep could  be a spurion which arises as a consequence of
spontaneous symmetry breaking at high energy by a field $\phi$, which is
communicated to the low energy theory by particles of mass $M$. Thus we
postulate that \ep$=\vev\phi/M$. Now in the effective theory supersymmetric
couplings in the superpotential can only depend on $\phi$ and not $\phi^*$. If
$\phi$ carries a PQ charge of -1, then the coupling  $(\vev\phi/M) \mu'\hu\hd$
is allowed, where we assume $\mu'$ is a parameter of order the supersymmetry
breaking scale. Couplings such as $(\vev\phi^*/M)\hdp\lambda'_{ij}   q_i \bar
d_j$ are however not allowed by supersymmetry. The coupling
$(\vev\phi^*/M)\hup\lambda''_{ij}   q_i \bar u_j$ is also  not allowed. There
could be some additional FCNC induced by $\hdp$ in the soft supersymmetry
breaking terms, but these should be acceptably small in any model which has
managed to solve the supersymmetric flavor problem \refs{\sfcncsol, \nir}.

Large \tb\ is interesting in the context of flavor models, which explain the
structure of the quark mass matrices  by approximate symmetries,
because it changes the charge assignments required in order to generate the
correct texture for the mass matrices.  For example,  we could assume that
there is another approximate U(1), which is explicitly broken by a spurion of
size $\alpha\sim0.2$ carrying charge -1.  Then for example we could
assign   charges (3,2,0) to the left handed quarks, (1,0,0) to the right
handed down type quarks, and (4,1,0) to  the right handed up type quarks.
Now the quark mass matrices naturally have the form\eqn\qumass{
\eqalign{
M_{\rm up}=&175\ \GeV\pmatrix{
\alpha^7&\alpha^4&\alpha^3\cr
\alpha^6&\alpha^3&\alpha^2\cr
\alpha^4&\alpha&1\cr}
\cr
M_{\rm down}=&{175\ \GeV\over\tan\beta}\pmatrix{
\alpha^4&\alpha^3&\alpha^3\cr
\alpha^3&\alpha^2&\alpha^2\cr
\alpha&1&1} \ ,\cr
}  } which gives the correct pattern of masses and
mixing angles for the quarks, up to factors of two.

   \newsec{Conclusions}

Our approach is orthogonal to that of previous authors. Rather
than confining our attention to a two doublet model, we
considered the question of whether it is possible to {\it naturally}
generate a large value for \tb, and if so, what if any are the
generic predictions of such a model. We have shown that it
is simple to make such a model at the expense of complicating the
Higgs sector;  we have shown furthermore that it is impossible to do
so without   additional charged fields. From this model,
we see that although the
minimal Higgs sector model with large \tb\ requires
the presence of additional light fields, there are  not necessarily additional
{\it light } fields present in a
nonminimal model;  the minimal model with large \tb\ can be made
natural  only at the expense of requiring light charginos.  Of course, the
minimal model, which contains an unnatural fine tuning of order 1/\tb\ in
order to be phenomenologically acceptable,  is very predictive
\refs{\als, \hall}. In nonminimal models, which can be natural, none
of these predictions necessarily apply.
\bigskip
{\bf Acknowledgements:}
We are grateful to the CERN theory division for their hospitality.
We thank David Kaplan for useful conversations.

\listrefs
\end